# π Berry phase and Zeeman splitting of TaP probed by high field magnetotransport measurements

J. Hu[1], J.Y. Liu[1], D. Graf[2], S.M.A Radmanesh[3], D.J. Adams[3], A. Chuang[1], Y. Wang[1], I. Chiorescu[2,3], J. Wei[1], L. Spinu[4] and Z.Q. Mao[1*]

[1]Department of physics and Engineering Physics, Tulane University, New Orleans, Louisiana 70118, USA

[2]National High Magnetic Field Laboratory, Florida State University, Tallahassee, Florida 32310, USA

[3]Department of Physics, Florida State University, Tallahassee, Florida 32306, USA

[4]Advanced Materials Research Institute and Department of Physics, University of New Orleans, New Orleans, Louisiana 70148, USA

Abstract

The chiral anomaly-induced negative magnetoresistance and non-trivial Berry phase are two fundamental transport properties associated with the topological properties of Weyl fermions. In this work, we report the quantum transport of TaP single crystals in magnetic field up to 31T. Through the analyses of our magnetotransport data, we show TaP has the signatures of a Weyl state, including light effective quasiparticle masses, ultrahigh carrier mobility, as well as negative longitudinal magnetoresistance. Furthermore, we have generalized the Lifshitz-Kosevich formula for Shubnikov-de Haas (SdH) oscillations with multi-frequencies, and determined the Berry phase of π for multiple Fermi pockets in TaP through the direct fitting of the quantum oscillations. In high fields, we also probed signatures of Zeeman splitting, from which the Landé *g*-factor is extracted.

* zmao@tulane.edu

The recent breakthrough in the discovery of Dirac [1-6] and Weyl [7-13] fermions in semimetals provides opportunities to explore the exotic properties of relativistic fermions in condensed matter. In Dirac semimetals, the discrete band touching points near the Fermi level - the Dirac nodes - are protected from gap opening by crystalline symmetry. The linear energy band dispersion near Dirac nodes hosts Dirac fermions whose low energy physics can be described by Dirac equations. With lifting spin degeneracy by breaking either time-reversal or inversion symmetry, a Dirac node is expected to split to a pair of Weyl nodes with opposite chirality [7-8]. The conservation of chirality of such pairs of separated Weyl nodes in momentum space leads to their topological robustness against translational symmetry invariant perturbations [7-8]. Near Weyl nodes, electron behaves as Weyl fermions and results in exotic signatures in Weyl semimetal, such as surface Fermi arcs that connect Weyl nodes of opposite chirality [14]. Moreover, Weyl fermions in Weyl semimetals also manifest themselves with exotic signatures in electron transport, such as chiral anomaly-induced negative longitudinal magnetoresistance (Adler-Bell-Jackiw chiral anomaly) and non-trivial Berry phase.

The recently proposed Weyl semimetal phase in transition metal monopnictides TX (T=Ta/Nb, X=As, P) [7-8] has stimulated intensive interests. Unlike the previously proposed magnetic Weyl semimetals [14-15], Weyl nodes appearing in TX is due to broken inversion symmetry [7-8]. Discontinuous surface Fermi arcs has been observed in photoemission spectroscopy studies on these materials [9-12, 16], confirming the existence of Weyl nodes. In addition, the Adler-Bell-Jackiw chiral anomaly, which is reflected as negative longitudinal magnetoresistance (MR) [17-19], has also been observed in TaAs [20-22], TaP [23-24], NbAs [25], and NbP [26].

In additional to the chiral anomaly-induced negative MR, non-trivial Berry phase is another important characteristic which can be revealed in the transport measurements. Berry phase describes the additional geometrical phase factor acquired in the adiabatic evolution along a closed trajectory in the parameter space [27]. Such additional phase does not depends on the details of the temporal evolution and thus differs from the dynamical phase [27-28]. In condensed matter, the Berry phase is determined by the topological characteristics of the electron bands in the Brillouin zone [28-30]. A non-zero Berry phase reflects the existence of band touching point [28] such as Dirac nodes, and manifest itself in observable effects in quantum oscillations. The cyclotron motion (that is, closed trajectory in momentum space) of Dirac fermions under magnetic field $B$ induces Berry phase that changes the phase of quantum oscillations. Generally, through mapping the Landau level (LL) fan diagram ($n^{th}$ LL index *vs.* $1/B_n$, the inverse of the applied magnetic field), the Berry phase can be conveniently extracted from the intercept of the linear extrapolation of LL index to the zero of inverse field $1/B$. Experimentally, a Berry phase of $\pi$ arising from the linear band dispersion of a Dirac cone [29] has been probed from the Shubnikov–de Haas (SdH) oscillations in both two dimensional (*e.g.* graphene [31-32], topological insulators [33-34], and $SrMnBi_2$ [35]) and three dimensional (*e.g.* $Cd_3As_2$ [36-37]) Dirac fermion systems.

In Weyl semimetals, a similar linear band dispersion [7-12, 16] also generates a non-trivial Berry phase. However, unlike the well-established $\pi$ Berry phase in Dirac systems as mentioned above [31-37], the experimental determination of Berry phase using the LL fan diagram remains elusive for monopnictide Weyl semimetals. First, the identification of the integer LL indices is inconsistent among published works; both the resistivity minima [25] and maxima [20-22, 26] have been used to assign the integer LL indices. Secondly, while all of these reports claim a non-trivial or π Berry phase, the intercept values obtained from the LL index plot, which is used to derive Berry phase, are diverse: 0 [20-21] and -0.08 [22] for TaAs, 0.12 [25, 38] for NbAs, and 0.32 [26] for NbP. As for TaP, however, Berry phase remains unexplored in the earlier studies [23-24]. Indeed, the LL fan diagram technique may not be an efficient method to extract the precise value of Berry phase in these monopnictide Weyl semimetals. Unlike previously studied Dirac systems which display only single frequency in SdH oscillations [31-37], monopnictide Weyl semimetals exhibit quantum oscillations with multiple frequencies due to the existence of multiple Fermi pockets [7-11, 16]. Therefore the oscillation peaks may not accurately correspond to the LL indices due to wave superposition. In addition, if one oscillation frequency is close to another (which is commonly seen in these systems for some certain field orientation [23, 26, 38]), separating the individual peak requires high magnetic field which was not achieved in previous low field studies [20-22, 25-26].

In order to address these issues, we conducted systematic high field magnetotransport measurements on TaP single crystals up to 31T. Our TaP crystal used in this study has superior quality, as demonstrated in the extremely large MR ($1\times10^6$ % at 1.6K and 31T, 450% at 300K and 14T) and ultra-high mobility ($1.2\times10^6$ cm$^2$/V s at 1.6K). The availability of such high quality crystals allow us to probe intrinsic quantum transport properties of TaP. Instead of using the LL fan diagram technique to extract the Berry phase, we have generalized the Lifshitz-Kosevich (LK) formula [39-41] and fitted the multi-frequency SdH oscillations in TaP. This approach is capable of revealing Berry phases for multiple Fermi pockets. Through this approach, we find that electrons from multiple Fermi pockets have π Berry phases accumulated along their cyclotron orbits, which agrees well with the nature of Weyl fermions. In high fields, we also observed Zeeman spin splitting, which enables us to extract the Landé *g*-factor for TaP for the first time.

The high quality TaP single crystals were synthesized using a chemical vapor transport technique. The stoichiometric mixture of Ta and P powder was sealed in a quartz tube with iodine as transport agent (20 mg/cm$^3$). Convex polytopes-like single crystals with metallic luster as large as 4mm×2.5mm×2mm (Fig. 1a, inset) can be obtained via the vapor transport growth with a temperature gradient from 950 °C to 850 °C. The composition and structure of TaP single crystals were checked by X-ray diffraction and Energy-dispersive X-ray spectrometer. Samples for transport measurements were polished to be plate-like with (100) surface. The magnetotransport measurements are performed using a Physics Property Measurement System and the 31 T resistive magnet at National High Magnetic Field Laboratory (NHMFL) in Tallahassee.

Figure 1 shows the magnetotransport properties of TaP with magnetic field along the crystallographic [100] axis and perpendicular to the current direction. The zero-field resistivity of TaP displays metallic behavior, with residual resistivity ~ 1.2 μΩ cm at 2K and the residual resistivity ratio (RRR) $\rho(300K)/\rho(2K)$ ~ 65, implying high quality of our single crystals. Applying magnetic field perpendicular to the current direction ($B\perp I$) induces extremely large MR and changes the temperature dependence of resistivity dramatically. A resistivity upturns appears when field $B > 0.1T$, which becomes more significant with increasing magnetic field and develops to an insulting-like behavior in full temperature range when $B > 3T$ (Fig. 1a). Such "metal - to - insulator-like" evolution driven by magnetic field is also observed in WTe$_2$ [42] and consistent with pervious observations in TaP [24] and other Weyl semimetals [20-22, 25, 43-44]. The extremely large MR can be better visualized in the field dependence of resistivity presented in Fig. 1b. At $T$ = 1.6K, the normalized MR $\Delta\rho_{xx}/\rho_0$ (=[$\rho_{xx}(B)$-$\rho_{xx}(B=0)$]/$\rho_{xx}(B=0)$) reaches $3\times10^5$% at 9T, larger than pervious observations [23]. Further increasing field lead MR to reach $1\times10^6$% at 31T without obvious sign of saturation, and SdH oscillations become visible above 9T. Even at room temperature, giant MR as large as 300% can be observed at $B$=9T (Fig. 1b, inset). It is surprising to observe such huge MR in a sample with RRR~65, given that large MR is usually accompanied with RRR of a few hundreds or even thousands [42, 45-46]. In fact, large MR with RRR less than 100 is commonly observed in monopnictide Weyl semimetals [20-26, 38, 43-44], in contrast with Dirac semimetal $Cd_3As_2$ in which MR and RRR are more closely related [33-34, 47]. In addition to the large magnitude, a linear field dependence for MR (MR$\propto B$) is also widely reported in monopnictide Weyl semimetals [20-22, 24-26, 43]. In our TaP single crystals, following a conventional quadratic dependence in low fields, a similar linear MR is also observed from $B$ = 3T up to 16T. Nevertheless, in the high field region which was not well explored in the previous studies, a power law dependence ~$B^{0.65}$ is found to better describe the MR data, suggesting that additional transport mechanisms may occur in high field.

To further characterize the electronic properties of TaP, we have measured the Hall resistivity $\rho_{xy}$ up to high fields. As shown in Fig. 1c, at low temperatures, the negative slop of linear field dependence of $\rho_{xy}(B)$ in high field range indicates that the transport is dominated by electrons; whereas the appearance of the nonlinear curvature in low field region implies the involvement of hole-type carriers. Indeed, both the theoretical calculations [7-8] and photoemission studies [9-12, 16] reveal both electron- and hole- Fermi surfaces in monopnictide Weyl semimetals. With increasing temperature, the hole-type carriers gradually dominate the transport above 150K, as indicated by the positive slop of $\rho_{xy}$(B) at high field (Fig. 1c, inset). Such evolution leads to the sign change of Hall coefficient $R_H$ that was derived from the linear dependence of $\rho_{xy}(B)$ at high field (Fig. 1d), consistent with the previous reports on monopnictide Weyl semimetals [20-26, 43-44].

From two-band model fitting [48], we have obtained ultra-high mobility $\mu_e \sim 1.8 \times 10^5$ cm$^3$/V s and $\mu_h \sim 2.5 \times 10^6$ cm$^3$/V s at 1.6K for electrons and holes respectively, as well as a nearly ideal electron and hole compensation with $n_e \sim 11.53 \times 10^{19}$ cm$^{-3}$ and $n_h \sim 11.50 \times 10^{19}$ cm$^{-3}$. Although the two-band model should be more reliable in such a multiband system, the possible low field anomalous Hall effect in Weyl semimetal [49-50] may lead the low field behavior of $\rho_{xy}(B)$ to deviate from the two-band model. Therefore we have also estimated the single band mobility using the Hall coefficient obtained at high field, which is approximately valid when one band dominates (*e.g.* at the lowest temperature, 1.6K). The obtained mobility is $\mu = R_H/\rho_{xx}(B=0) = 2.1 \times 10^6$ cm$^3$/Vs, consistent with those obtained from the two-band model fitting.

The observed high mobility is larger than that in previous observations in TaP (e.g., $9 \times 10^5$ cm$^3$/Vs in Ref. [23] and $2 \times 10^5$ cm$^3$/Vs in Ref. [24]), and comparable to that of NbP - the largest mobility seen in the monopnictide Weyl semimetal family [26, 44]. Such high mobility may be responsible for the rapid increase of resistivity with magnetic field as described by $\rho \sim (\mu B)^2$, while the nearly ideal electron-hole compensation prevents the saturation of MR in high fields. However, this simple classical model fails to interpret the linear and $\sim B^{0.65}$ dependence of MR at high fields, suggesting the involvement of other mechanisms such as quantum linear MR [51] or plausible broken topological protection under magnetic field [52].

In Weyl semimetals, the pair of Weyl nodes acts as source and drain of Berry flux, leading to a non-zero Berry curvature $\mathbf{\Omega}_p$ and causing an additional topological contribution to the Weyl fermions dynamics, which is proportional to the product of electric and magnetic field,

*i.e.*, $\propto(\boldsymbol{E}\cdot\boldsymbol{B})\Omega_p$ [19]. As a consequence, non-orthogonal electric and magnetic field ($\boldsymbol{E}\cdot\boldsymbol{B}\neq 0$) can lead to charge transfer between two Weyl nodes with opposite chirality. Such violation of the chiral charge conservation is known as the Adler-Bell-Jackiw anomaly or chiral anomaly [17-19], and results in negative MR that can take place in semiclassical regime [18-19]. As shown in Figs. 2a and 2c, with magnetic field aligned parallel to the current direction (*B*//*I*), we observed negative longitudinal MR above *B*=0.5T at 1.6K, which can be quickly suppressed when the field orientation is deviated from the current (Fig. 2b), in agreement with the pervious observations in Weyl semimetals [20-25]. The positive MR below 0.5T may be ascribed to weak antilocalization due to strong spin-orbit coupling. With increasing field above 4T, the negative MR turns to be positive, which roughly increases with $B^2$ and is accompanied by strong SdH oscillations. This suggests a dominating classical orbital MR at high fields which is probably induced by non-ideal parallel alignment between *B* and *I*, or Fermi surface anisotropy [53]. With the suppression of the classic MR by rising temperature, negative MR becomes more prominent and extends to high field up to 31T, without any signature of saturation (see the bottom panel in Fig. 2a). Such results are consistent with the earlier low field experiments for TaP [23], and provide the first direct observation of significant chiral anomaly contribution to transport at high fields. The strong negative MR at high temperatures suggests that Weyl fermions make more significant contributions in TaP as compared to other monopnictide Weyl semimetals. This is consistent with the recent photoemission study which found that the four Weyl points locate at the Fermi energy [12].

Generally, in the quantum limit where all electrons are condensed to the lowest Landau level (LL), only the LL=0 chiral branch is occupied and the intra-node scattering is prohibited. Therefore, the longitudinal current (*E*//*B*) can be relaxed only via inter-node scattering, causing the enhanced negative MR [17-18]. Indeed, a previous study has found that the negative MR in TaP is enhanced upon cooling and persists up to 14T [23], in sharp contrast with our observations of the high field (> 4T) positive MR at 1.6K (Fig. 2a). This can be understood in terms of the ultra-high mobility of our TaP single crystals (an order of magnitude larger than that in Ref. [23]), which gives rise to much stronger classical MR since MR$\propto\mu^2 B^2$.

More quantitative information of Weyl fermions transport in TaP can be extracted through their $B^2$-dependence of magnetoconductivity (MC) [18-19]. The total conductivity can be written as $\sigma = \sigma_0(1+C_w B^2)$ in parallel field, in which $\sigma_0$ is the normal conductivity and the chiral anomaly contribution is thus to be $\sigma_0 C_w B^2$ [19]. In our sample, the presence of weak antilocalization leads to correction to normal conductivity, *i.e.*, $\sigma_0 + a\sqrt{B}$ [54]. With the consideration of the additional classic positive MR $\propto B^2$ [54], the MC of TaP can be written as:

$$\sigma(B) = (\sigma_0 + a\sqrt{B})(1+C_w B^2) + \frac{1}{\rho_1 + A_1 B^2} + \frac{1}{\rho_2 + A_2 B^2} \quad (1)$$

Similar to the previous reports, two classic MR terms are necessary to obtain the satisfactory fit [21, 23]. As shown in Fig. 2c, the above equation fits the chiral anomaly induced positive MC (negative MR) very well for all measured temperatures, as shown by the black solid fitting curves in Fig. 2c. However, the high field negative MC (positive MR) at low temperatures cannot be simultaneously described by Eq. 1. This can be understood in terms of approaching quantum limit at low temperatures, where the contribution from chiral anomaly becomes linearly dependent on $B$ [17-19].

From the fitting we have extracted the chiral anomaly contribution $\sigma_w = \sigma_0 C_w B^2$ [19] at various temperatures. In principle, $C_w$ is $T$-independent and the temperature variation of $\sigma_w$ results from $\sigma_0(T)$ ($\propto \mu(T)$). As expected, $C_W$ displays very weak temperature dependence, as shown in the inset of Fig. 2c.

In addition to the chiral anomaly induced negative MR, the non-trivial Berry phase accumulated in the cyclotron motion of Weyl fermions is another fundamental topological property of Weyl semimetals, since that the Berry phase takes the value of 0 and π for parabolic and linear band dispersions, respectively [29]. The effect of this additional phase factor in SdH oscillation can be described by the Lifshitz-Kosevich (LK) formula which is developed for 3D system with arbitrary band dispersions [39-41]:

$$\frac{\Delta\rho}{\rho(B=0)} = \frac{5}{2}\left(\frac{B}{2F}\right)^{1/2} \frac{2\pi^2 k_B T m^* / \hbar eB}{\sinh(2\pi^2 k_B T m^* / \hbar eB)} e^{2\pi^2 k_B T_D m^* / \hbar eB} \cos[2\pi(\frac{F}{B} + \gamma - \delta + \phi)] \quad (2)$$

In Eq. 2, the hyperbolic and exponential terms describe the temperature and field damping of oscillation amplitude, which are determined by effective mass $m^*$ and Dingle temperature $T_D$. The oscillation frequency is described by the cosine term that contains a phase factor $\gamma - \delta + \phi$, in which $\gamma = \frac{1}{2} - \frac{\phi_B}{2\pi}$ and $\phi_B$ is Berry phase. The phase shift $\delta$ is determined by the dimensionality of Fermi surface and takes values 0 for 2D and ±1/8 for 3D cases. The additional phase factor $\phi$, which is missing in the original LK formula [39-41], is due to the relative phase between conductivity and resistivity oscillations, i.e. $\Delta\sigma$ and $\Delta\rho$ [57]. Given that $\sigma_{xx} = \rho_{xx}/(\rho_{xx}^2 + \rho_{xy}^2)$, when the longitudinal resistivity $\rho_{xx} \gg$ transverse (Hall) resistivity $\rho_{xy}$, the oscillation component of $\sigma_{xx}$ obtained by taking the derivative is $\Delta\sigma_{xx} \approx \Delta\frac{1}{\rho_{xx}} = -\frac{\Delta\rho_{xx}}{\rho_{xx}^2}$, indicating completely out-of-phase for $\Delta\sigma$ and $\Delta\rho$. In contrary, when $\rho_{xx} \ll \rho_{xy}$, $\Delta\sigma$ and $\Delta\rho$ are in phase since

$\Delta\sigma_{xx} \approx \Delta(\frac{\rho_{xx}}{\rho_{xy}^2}) = \frac{1}{\rho_{xy}^2}\Delta\rho_{xx} - \frac{2\rho_{xx}}{\rho_{xy}^3}\Delta\rho_{xy} \approx \frac{1}{\rho_{xy}^2}\Delta\rho_{xx}$. Therefore, when $\sigma_{xx}$ is maximized with lifting LL to Fermi energy, $\rho_{xx}$ can display either minimum ($\rho_{xx} \gg \rho_{xy}$) or maximum ($\rho_{xx} \ll \rho_{xy}$), leading to an additional phase factor $\phi = 1/2$ or $0$, respectively, in the LK formula.

Based on the above analysis, resistivity exhibits minimum ($\rho_{xx} \gg \rho_{xy}$) or maximum ($\rho_{xx} \ll \rho_{xy}$) with successive filling of the $n^{th}$ LL. Therefore from Eq. 2 we can obtain the Lifshitz-Onsager quantization rule: $\frac{F}{B_n} + \gamma - \delta = n$ (when $\rho_{xx} \ll \rho_{xy}$) or $n + \frac{1}{2}$ (when $\rho_{xx} \gg \rho_{xy}$). Such linear relation between $1/B_n$ and $n$ allows the extraction of $\gamma - \delta$ and consequently the Berry phase through the intercept of the LL fan diagram, as seen in a variety of Dirac systems [31-37]. However, the LL fan diagram approach is less efficient in monopnictide Weyl semimetals due to multiple frequencies involved in oscillations as discussed above. In Fig. 3a we show the oscillatory component of resistivity $\Delta\rho$ for $B\perp I$ at various temperatures obtained by subtracting the smooth MR background ($\Delta\rho = \rho - \rho_{br}$). A clear splitting can be observed at $0.04\text{T}^{-1}$ ($B=25\text{T}$), which can be ascribed to Zeeman splitting and will be discussed later. Except for this feature, $\Delta\rho$ is, however, still not exactly periodic in $1/B$. In fact, the fast Fourier transfer (FFT) of $\Delta\rho$ (Fig. 3b, inset) yields two major frequencies at $F_\alpha = 110\text{T}$ and $F_\beta = 150\text{T}$ with comparable amplitude (the high field splitting data is not included for FFT). The superposition of two oscillation waves could cause broadening and even shift of the oscillation extrema, making it is difficult to assign LL index. Similar situation also occurs for the case of $B//I$ (Figs. 4a and 4b), in which major frequencies of $F_\gamma = 18\text{T}$ and $F_\delta = 44\text{T}$ can be resolved via FFT transform.

Similar multi-frequency problem also occurs in other monopnictide Weyl semimetals due to their multiple Fermi pockets. Most of earlier works focused on the "major" oscillation extrema and found the corresponding Berry phase for one Fermi pocket [20, 22, 25-26]. An approach is to use the 2$^{nd}$ derivative of $\Delta\rho$ to separate oscillation peaks for different frequencies [38]. A variety of quite different intercepts in LL fan diagram has been obtained in those works as mentioned above [20, 22, 25-26, 38]. Such inconsistence may be associated with sample variation or, more likely, the uncertainty in identifying $1/B_n$ due to the superposition of oscillation peaks.

In TaP, Berry phase has not been explored in previous magnetotransport studies [23-24]. In order to extract the accurate Berry phase in TaP, instead of LL fan diagram, we attempted to use the LK formula (Eq. 1) to fit the multi-frequency SdH oscillations directly. Given that we have observed two major frequencies for both $B \perp I$ and $B//I$, we can reasonably assume that the total resistivity $\rho$ due to two Fermi pockets follows $1/\rho=(1/\rho_1+1/\rho_2)$. Taking the differential we can obtain $\Delta\rho = \frac{\rho_2^2}{\rho_1^2+\rho_2^2}\Delta\rho_1 + \frac{\rho_1^2}{\rho_1^2+\rho_2^2}\Delta\rho_2$, indicating that the resistivity oscillations are additive and the LK formula (Eq. 2) can be easily generalized for multi-frequency oscillations by linear superposition. To reduce the fitting parameters, the effective mass $m^*$ can be first extracted through the temperature dependence of the amplitude of FFT for $\Delta\rho/\rho_0$, using the thermal damping term in the LK formula, *i.e.* $\frac{2\pi^2 k_B T m^*/\hbar e\overline{B}}{\sinh(2\pi^2 k_B T m^*/\hbar e\overline{B})}$ with $1/\overline{B}$ being the average inverse field. The values of $m^*$ are considerably smaller for $\gamma$ and $\delta$ Fermi pockets ($m_\gamma=0.04m_0$ and $m_\delta=0.08\ m_0$, $m_0$ is the free electron mass, Fig. 4b) than those of $\alpha$ and $\beta$ Fermi pockets ($m_\alpha=0.26m_0$ and $m_\beta=0.24m_0$, Fig. 3b), consistent with the previous results [23].

To avoid the influence from Zeeman splitting, during the fitting using the LK formula, we have fitted the SdH oscillations at 10K where the Zeeman effect is minimized by the thermal broadening of LL, as well as the low temperature ($T$=1.6K) data with the high field splitting being excluded. As shown in Figs. 3c and 4c, the two-band LK formula reproduces the resistivity oscillations very well. The phase shift $\gamma - \delta + \phi$ extracted from fitting are 0.42 and 0.41 for $\alpha$ and $\beta$ Fermi pockets when $B \perp I$, and 0.12 and 0.67 for $\delta$ and $\gamma$ Fermi pockets when $B//I$. Taking $\gamma = \frac{1}{2} - \frac{\phi_B}{2\pi}$ and $\delta = \pm 1/8$ expected for 3D Fermi surface, a Berry phase of $\pi$ lead the phase shift $\gamma - \delta + \phi$ (which is also the intercept of the LL fan diagram, if the LL indices can be accurately assigned) to be $\phi \pm 1/8$, which is 0.375 or 0.625 when $\rho_{xx} \gg \rho_{xy}$, and $\pm 1/8$ when $\rho_{xx} \ll \rho_{xy}$. Given that $\rho_{xx} \gg \rho_{xy}$ for $B \perp I$ and $B//I$ ($\rho_{xy}$=0 for $B//I$) in our sample, the obtained phase shift from two-band LK fitting implies a $\pi$ Berry phase for $\alpha$, $\beta$, and $\gamma$ Fermi pockets (in particular, 0.91$\pi$, 0.93$\pi$, and 0.91$\pi$ for $\alpha$, $\beta$, and $\gamma$ Fermi pockets), as well as a trivial Berry phase near zero for the $\delta$ Fermi pocket. The determination of nearly $\pi$ Berry phase is consistent with the linear band dispersion of Weyl fermions in TaP.

In addition to the Berry phase, with high field measurements we are able to determine the Landé $g$-factor through Zeeman splitting for the first time. As stated above, we have observed peak splitting in high field oscillations (0.04T$^{-1}$ for $B \perp I$ and 0.047 T$^{-1}$ for $B//I$, see Figs. 3a and 4a). With raising temperature above 10K, the splitting gradually merges to single peak, implying an origin of the Zeeman effect. Such splitting becomes more significant in the oscillation of Hall resistivity $\Delta\rho_{xy}$, and also weakens with raising temperature, as shown in Fig. 3d. Unfortunately, we are unable to separate electrons from different Fermi surfaces in SdH oscillations, but an average $g$-factor of 2-2.9 can be obtained for $\alpha$ and/or $\beta$ Fermi pockets, and 5.5 – 6.7 for $\delta$ and/or $\gamma$ Fermi pockets.

In summary, we have preformed systematic high field magnetotransport measurements on high quality TaP single crystals, which display extremely large magnetoresistance and ultra-high mobility. We have found the chiral anomaly-induced negative MR can extend up to 31T at high temperatures, implying significant involvement of Weyl fermions in transport. More importantly, using the generalized two band LK formula, we have successfully determined the non-trival Berry phase very close to $\pi$ for multiple Fermi pockets in TaP. Such results not only reveal the unexplored fundamental topological properties of TaP, but also provide an effective approach that can be generalized to other Weyl semimetals.

Note: In the preparation of the manuscript, we were aware of a high field measurement which revealed quantum phase transition in TaP [58], and a low field magnetotrasport work which revealed a non-trivial Berry phase for a single Fermi pocket using LL fan diagram [59].

Acknowledgement

The work at Tulane is supported by the DOE under Grant No.DE-SC0012432 (support for materials, travel to NHMFL and personnel) and the National Science Foundation under the NSF EPSCoR Cooperative Agreement No. EPS-1003897 with additional support from the Louisiana Board of Regents (support for personnel). The work at UNO is supported by the NSF under the NSF EPSCoR Cooperative Agreement No. EPS-1003897 with additional support from the Louisiana Board of Regents. The work at FSU and at the National High Magnetic Field Laboratory, is supported by the NSF grant No. DMR-1206267, the NSF Cooperative Agreement

No. DMR-1157490 and the State of Florida. DG also acknowledges support from grant DOE DE-NA0001979.

**Figure 1**

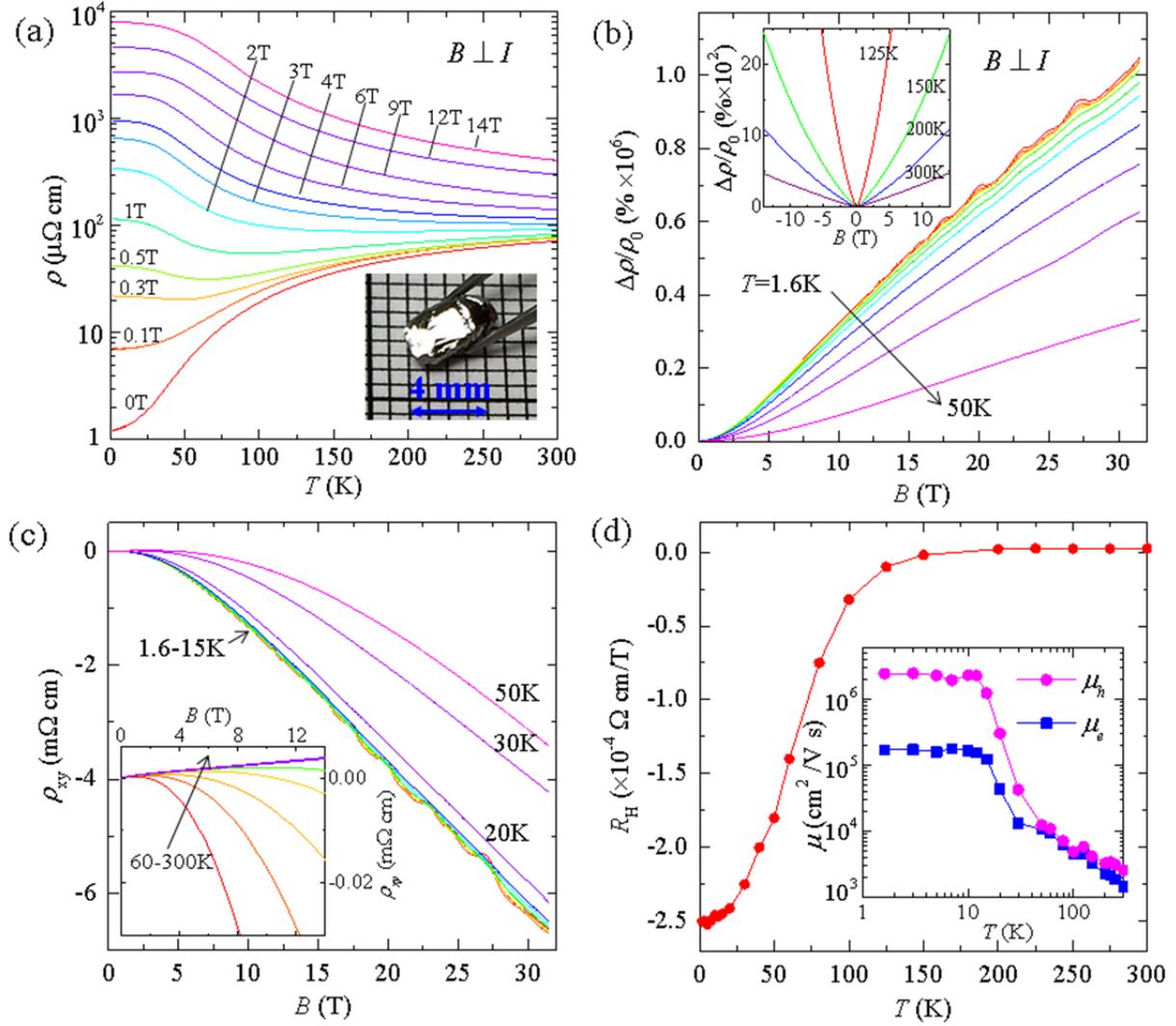

FIG. 1. Transport properties of TaP. (a) Temperature dependence of resistivity under various magnetic field B from 0 to 14T. The magnetic field is applied perpendicual to the current direction. Inset: image of a large TaP crystal. (b) Magnetoresistance $\Delta\rho/\rho_0=[\rho(B) - \rho(B=0)]/\rho(B=0)$ of TaP at various temperatures $T$. From bottom to top: $T$ = 50, 40, 30, 20, 15, 12, 10, 7, 5, 3, and 1.6K. Inset: Magnetoresistance at high temperatures. From bottom to top: T =300, 200, 150, and 100K (c) Hall resistivity $\rho_{xy}$ at various temperatures. The inset shows the $\rho_{xy}$ at higer temperature from 300K to 60K (d) Temperature dependence the Hall coefficient. The inset shows the mobility extracted from the two-band model fitting (see text).

**Figure 2**

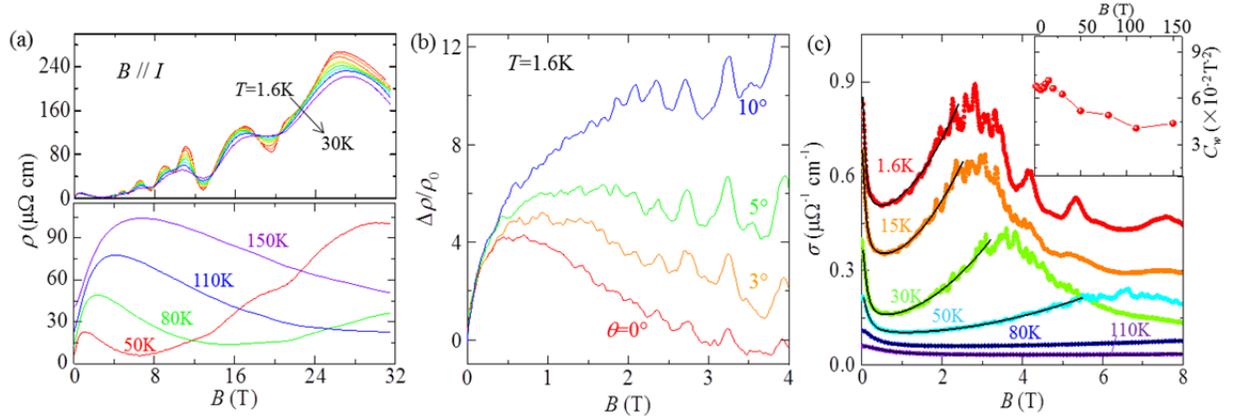

FIG. 2. Chiral anomaly-induced negative MR in TaP. (a) and (b) The longitudinal MR at (a) 1.6-30K and (b) 50-150K. (b) Angular dependence of longitudinal MR at 1.6K. (c) Fitting of the magnetoconductivity $\sigma(B)$ at different temperatures using Eq. 1 (see text). Data for different temperatures has been shifted for charlity. The inset shows the temperature dependence of Cw, which show very weak $T$ dependence.

**Figure 3**

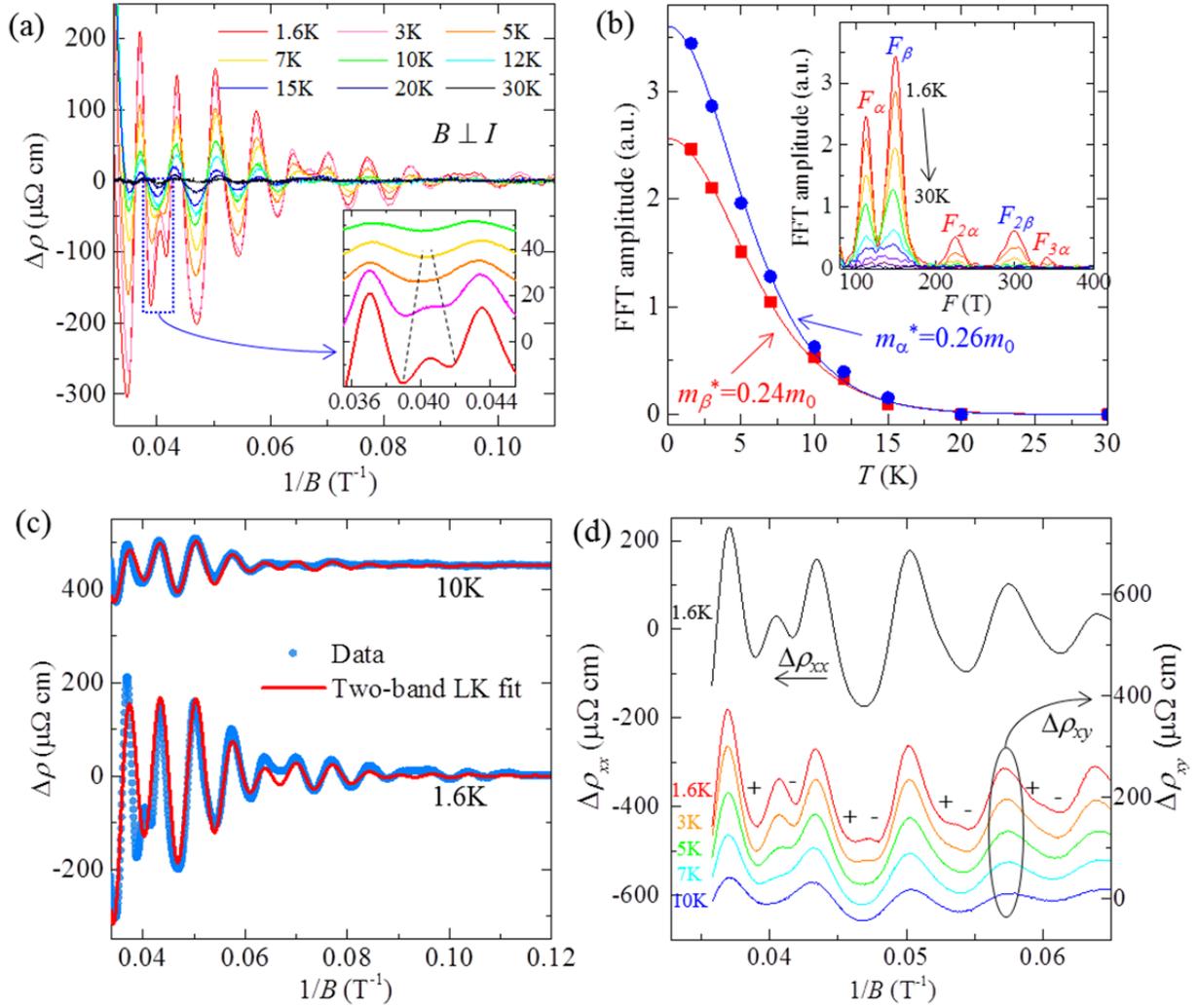

FIG. 3. SdH oscillation when $B \perp I$. (a) The oscillation component of resistivity at different temperatures, obtained by $\Delta\rho=\rho-\rho_{br}$. The inset shows the Zeeman splitting of an oscillation peak at high field, which gradually merges with raising temperature. Data at different temperatures has been shifted for clarity. (b) The temperature dependence of FFT amplitude for $\Delta\rho/\rho(B=0)$. The solid lines show the effective mass fitting according to the temperature damping term of the Lifshitz-Kosevich formula. The inset shows the FFT for $\Delta\rho/\rho(B=0)$ at various temperatures. (c) Fitting of SdH oscillation at 1.6K and 10K. The 10K data has been shifted for clarity (d) Zeeman splitting of longitudinal ($\Delta\rho_{xx}$) and Hall resistivity ($\Delta\rho_{xy}$). $\Delta\rho_{xy}$ at different temperatures has been shifted for clarity.

**Figure 4**

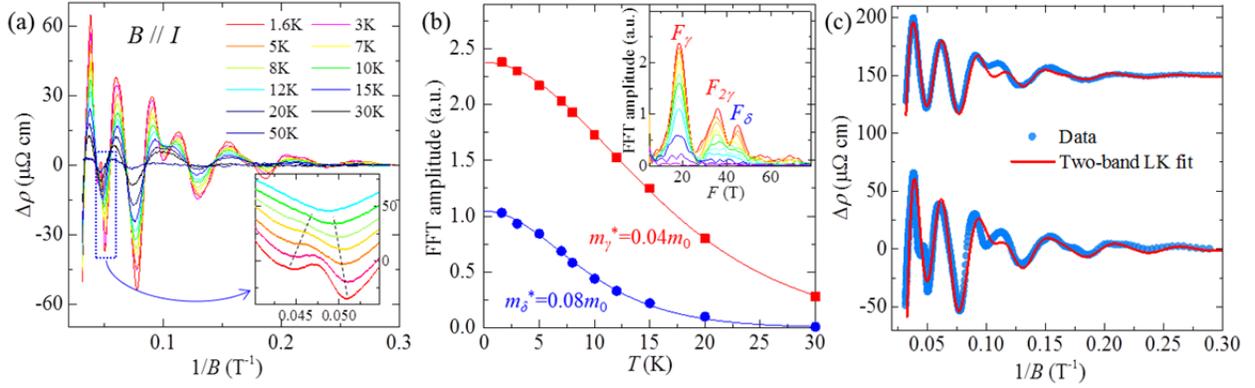

FIG. 4. SdH oscillation when $B//I$. (a) The oscillation component of resistivity $\Delta\rho=\rho-\rho_{br}$ at different temperatures. The inset shows the Zeeman splitting of a oscillation peak at high field, which gradually merges with raising temperature. Data at different temperatures has been shifted for clarity. (b) The temperature dependence of FFT amplitude for $\Delta\rho/\rho(B=0)$. The solid lines show the effective mass fitting according to the temperature damping term of the Lifshitz-Kosevich formula. The inset shows the FFT for $\Delta\rho/\rho(B=0)$ at various temperatures (1.6-50K). (c) Fitting of SdH oscillation at 1.6K and 10K. The 10K data has been shifted for clarity.